\pdfoutput=1
\documentclass[prb,amssymb,twocolumn,superscriptaddress]{revtex4-1}

\usepackage{lineno}
\usepackage{graphics}
\usepackage{mathtools}
\usepackage{float}
\usepackage{amsmath}

\begin{document}
\title{Superconductivity and Frozen Electronic States at the (111) LaAlO$_3$/SrTiO$_3$ Interface}

\author{S. Davis}
\email[]{samueldavis2016@u.northwestern.edu}

\affiliation{Graduate Program in Applied Physics and Department of Physics and Astronomy, Northwestern University, 2145 Sheridan Road, Evanston, IL 60208, USA}
\author{Z. Huang}
\author{K. Han}

\affiliation{NUSNNI-Nanocore, National University of Singapore 117411, Singapore}  
\affiliation{Department of Physics, National University of Singapore 117551, Singapore } 
\author{Ariando}
\affiliation{NUSNNI-Nanocore, National University of Singapore 117411, Singapore}  
\affiliation{Department of Physics, National University of Singapore 117551, Singapore } 
\affiliation{NUS Graduate School for Integrative Sciences \& Engineering, National University of Singapore 117456, Singapore}

\author{T. Venkatesan}
\affiliation{NUS Graduate School for Integrative Sciences \& Engineering, National University of Singapore 117456, Singapore}
\affiliation{NUSNNI-NanoCore, National University of Singapore 117411, Singapore}
\affiliation{Department of Physics, National University of Singapore 117542, Singapore}
\affiliation{Department of Electrical and Computer Engineering, National University of Singapore 117576, Singapore}\affiliation{Department of Material Science and Engineering, National University of Singapore 117575, Singapore}

\author{V. Chandrasekhar}
\email[]{v-chandrasekhar@northwestern.edu}
\affiliation{Graduate Program in Applied Physics and Department of Physics and Astronomy, Northwestern University, 2145 Sheridan Road, Evanston, IL 60208, USA}

\date{\today}%

\maketitle


\textbf{In spite of Anderson's theorem\cite{and}, disorder is known to affect superconductivity in conventional $s$-wave superconductors\cite{bose,mats,abel,nitt, mond,mark,shim,kamla,deg,garland}.  In most superconductors, the degree of disorder is fixed during sample preparation.  Here we report measurements of the superconducting properties of the two-dimensional gas that forms at the interface between LaAlO$_3$ (LAO) and SrTiO$_3$ (STO) in the (111) crystal orientation, a system that permits \emph{in situ} tuning of carrier density and disorder by means of a back gate voltage $V_g$.  Like the (001) oriented LAO/STO interface\cite{thiel,reyren,caviglia,dikin}, superconductivity at the (111) LAO/STO interface can be tuned by $V_g$.  In contrast to the (001) interface, superconductivity in these (111) samples is anisotropic, being different along different interface crystal directions, consistent with the strong anisotropy already observed other transport properties at the (111) LAO/STO interface\cite{davis2,davis1}.  In addition, we find that the (111) interface samples ``remember'' the backgate voltage $V_F$ at which they are cooled at temperatures near the superconducting transition temperature $T_c$, even if $V_g$ is subsequently changed at lower temperatures.  The low energy scale and other characteristics of this memory effect ($<1$ K) distinguish it from charge-trapping effects previously observed in (001) interface samples.  }

For most conventional superconductors, the presence of disorder decreases the superconducting transition temperature, $T_c$\cite{bose,mats}.  While the exact reason for this is still a subject of investigation, it is thought that localization effects coupled with enhanced Coulomb interactions resulting from the short elastic mean free path give rise to an inhomogeneous superconducting state with large amplitude and phase variations, even though the disorder may be uniform\cite{mark,shim,kamla}.  In the limit of large disorder, the material may transition into an insulating state, the so-called superconductor-to-insulator transition (SIT)\cite{mark}.  In a few superconductors, increasing disorder increases $T_c$: perhaps the best-known example is Al, where the enhancement of $T_c$ with disorder can be as large as a factor of 5\cite{nitt,abel}.  In this case, the enhancement in $T_c$ is thought to be due to a modification of the electron-phonon interaction responsible for the attractive potential between quasiparticles.\cite{bose,deg,garland}

A SIT has been observed in the (001) LAO/STO interfaces\cite{thiel,reyren,caviglia,dikin}; unlike other thin films discussed above, the advantage of this two-dimensional (2D) system is that the system parameters can be changed by simply changing the voltage $V_g$ on a gate, enabling tuning of the SIT with an \emph{in situ} experimental handle.  As we show below, the (111) LAO/STO interface devices also show a superconducting transition that can be tuned by $V_g$.  Unlike the (001) LAO/STO devices, however, the superconducting characteristics are anisotropic, being different when measured along two mutually perpendicular crystal directions.  This behavior complements the strong anisotropy that we have reported recently in other transport properties of the (111) interface, including the longitudinal resistance, Hall effect and quantum capacitance \cite{davis2,davis1}:  as discussed there, the anisotropy is not due to the hexagonal symmetry of the (111) interface,  microstrucural effects or ferroelectric twin domains.  More remarkably, both the normal and superconducting properties at temperatures $T\sim T_c$ and below strongly depend on the gate voltage $V_F$ as the sample is cooled from $\sim 4$ K, even if $V_g$ is subsequently changed at lower temperatures.  This freezing effect is an indication of disorder that can be frozen in at very low temperatures in these structures by the application of a gate voltage, and which can be reversed by warming the sample to a freezing temperature $T_F\sim1$ K, and is distinctly different in energy scale and qualitative response to $V_g$ from the charge trapping behavior reported earlier in (001) LAO/STO interface samples\cite{caviglia,liu,bell,Biscaras}.
  
\begin{figure}[h!]
	\label{RvsT}
	\center{\includegraphics[width=7cm]{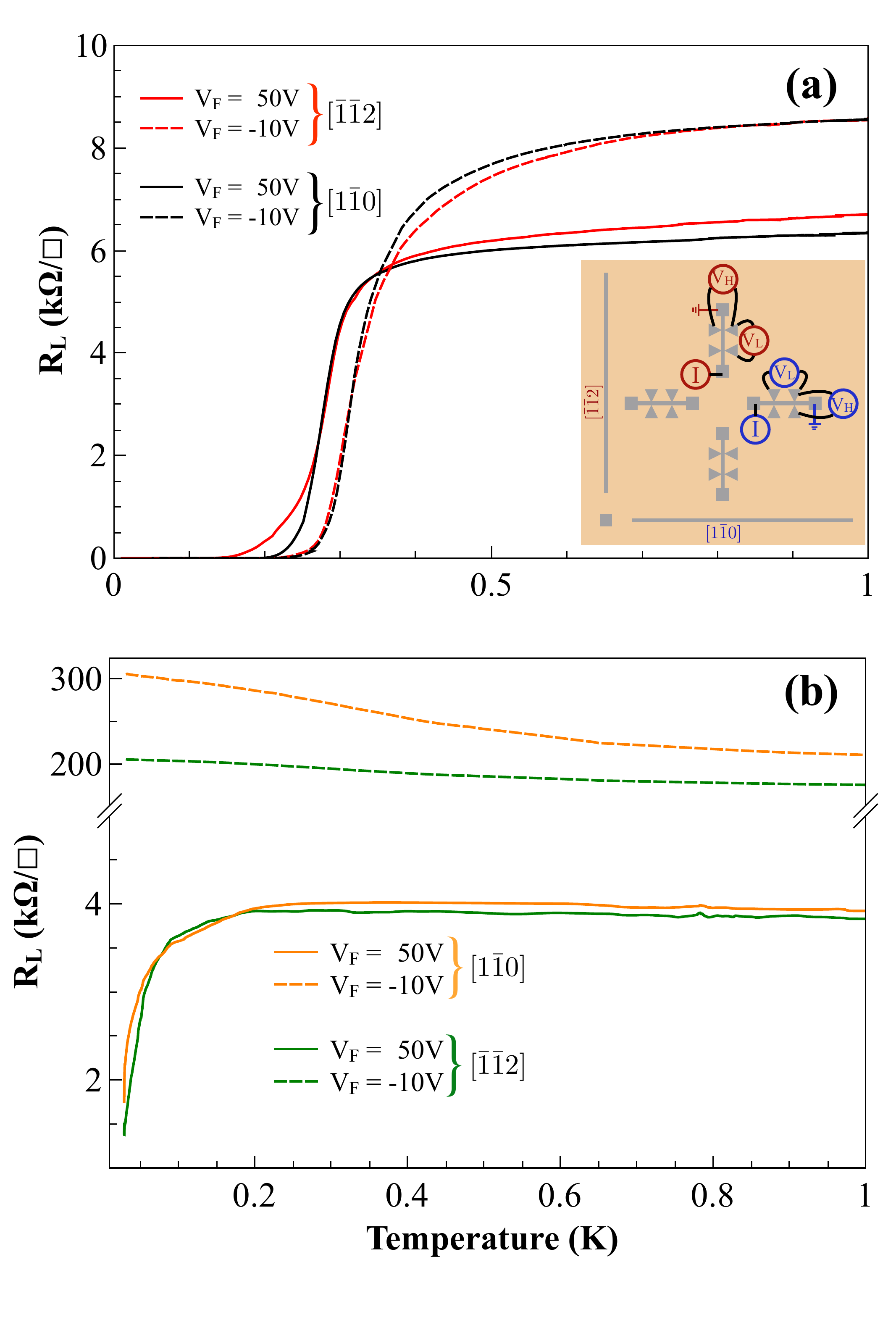}}
	\caption{\textbf{Resistance as a function of temperature with different freezing voltages $V_F$ along two crystal directions.}  The Ar/H$_2$ annealed samples \textbf{(a)} show full superconducting transitions.  While the normal state resistance $R_N$ is higher for $V_F=-10$ V, the superconducting transition temperature $T_c$ is also higher.  The O$_2$ annealed samples \textbf{(b)} do not go superconducting, but show hints of a superconducting transition for $V_F=50$ V, and insulating behavior for $V_F=-10$ V, with strong anisotropy between the the $[\bar{1}\bar{1}2]$ and $[1\bar{1}0]$ crystal directions in the latter.  Inset in (a) is a schematic of the Hall bars on each sample chip, which shows the measurement configuration.}
\end{figure}           

The heterostructures used in this work consisted of 20 monolayers (ML) of LAO grown epitaxially on Ti terminated (111) STO substrates by pulsed laser deposition.  Details of the film synthesis can be found in \emph{Methods}.  On each 5 mm$\times$5 mm chip, 4 Hall bars (600 $\mu$m $\times$ 100 $\mu$m) were fabricated using a combination of photolithography and Ar ion milling such that two of the Hall bars were oriented along the $[\bar{1}\bar{1}2]$ surface crystal direction, and the remaining two were oriented along the $[1\bar{1}0]$ crystal direction, as shown in the inset to Fig. 1(a).  Chips were then exposed to different post-growth annealing processes to change their global carrier densities.  Half of the devices were annealed in an O$_2$ atmosphere and the other half in a Ar/H$_2$ atmosphere (see \emph{Methods}).  Annealing in an O$_2$ environment reduces the number of oxygen vacancies and hence reduces the carrier concentration, while annealing in Ar/H$_2$ increases the oxygen vacancy concentration and hence increases the carrier concentration \cite{davis1}.  A total of 10 Hall bars on three different chips were measured; here we report data on the four Hall bars for which we have the most comprehensive data.

In (111) interface structures, the transport characteristics (longitudinal resistance, Hall coefficient, and quantum capacitance) are strongly anisotropic\cite{davis1,davis2}, being different along the $[\bar{1}\bar{1}2]$ and $[1\bar{1}0]$ directions, the anisotropy increasing with decreasing oxygen vacancy concentration and more negative $V_G$\cite{davis1}. As we show below, the superconducting characteristics are also anisotropic. Figure 1(a) shows the superconducting transition of a Ar/H$_2$ annealed sample with back gate `freezing' voltages $V_F=-10$ V and $V_F=50$ V applied as the sample was cooled through the transition. To maintain a uniform protocol for all data taken, $V_F$ was applied at a temperature of 4.4 K before cooling down to 30mK, unless otherwise noted. $V_G$ was then repeatedly cycled between +/- 100V to obtain reproducible curves (see Supplementary Information) before the data reported here were taken. 

The data discussed here are for two freezing voltages $V_F=-10$ V and $V_F=50$ V for which we have taken the most extensive data.  Data for other voltages (e.g., $V_F=0$ V) follow the expected trend (see Supplementary Information).  The Ar/H$_2$ annealed samples showed a superconducting transition, with $V_F$ affecting both the normal state resistance $R_N$ and the superconducting transition $T_c$.  As expected from our previous work \cite{davis2,davis1}, the Ar/H$_2$ annealed samples, with a larger number of oxygen vacancies, are not strongly anisotropic.  The O$_2$ annealed samples did not go fully superconducting: there is a hint of a superconducting transition for $V_F=50$ V, while insulating behavior is seen for $V_F=-10$ V, with clear anisotropy in resistance between the two crystal directions (a much stronger anisotropy is observed at lower gate voltages \cite{davis2,davis1}).  A similar dependence of the superconducting properties on gate voltage has been reported in (001) LAO/STO structures, and is ascribed to electrostatic doping of the electron gas.  We show below that part of the difference in the (111) LAO/STO devices is due to disordered trapped states frozen in by the application of a gate voltage at higher temperatures.

\begin{figure*}[!]
\label{Ar-dVdI}
\center{\includegraphics[width=14.5cm]{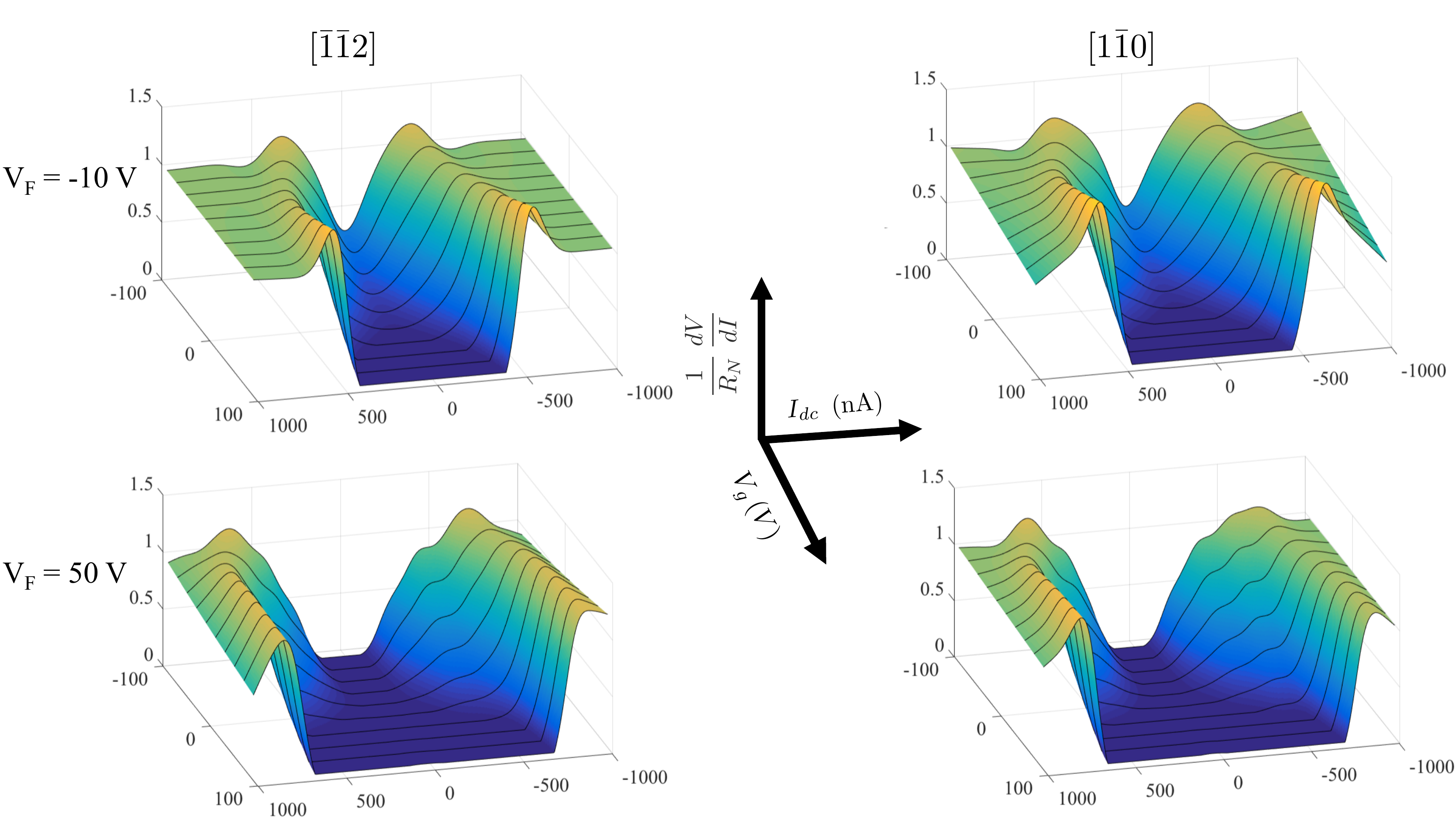}}
\caption{\textbf{Differential resistance of the Ar/H$_2$ annealed samples at 30 mK, for $V_F=-10$ V and $V_F=50$ V}.  
The critical current $I_c$ varies with $V_g$, a consequence of electrostatic doping by the gate, but is also a function of the freezing voltage $V_F$ applied while cooling through the superconducting transition:  in general, $I_c$ is larger for $V_F=50$ V in comparison to $V_F=-10$ V.  For the Ar/H$_2$ annealed devices, there is relatively little anisotropy between the two crystal directions.
Due to the large variation in $R_N$ with $V_g$, the curves here and in Fig. 3 are normalized to the value of $dV/dI$ at $\pm 1$ $\mu$A.}
\end{figure*} 

The difference between electrostatic doping and frozen disorder can be seen clearly in the nonlinear differential resistance $dV/dI$ as a function of dc current $I_{dc}$ and gate voltage $V_g$ at temperatures far below $T_c$, shown in Fig. 2 for the Ar/H$_2$ annealed samples.  As expected,the critical current $I_c$ varies with $V_g$, being in general smaller for more negative $V_g$.  This is the effect of electrostatic doping with $V_g$, similar to that observed previously in the (001) LAO/STO samples\cite{dikin}.  However, the most striking feature of this data is that the current-voltage characteristics also depend on the gate voltage $V_F$ at which the samples were cooled, even though $V_g$ is changed over the same range.  The difference is persistent:  as noted above, data were taken after the gate voltage had been swept repeatedly between $\pm$100 V at 30 mK; the sample remembers the voltage $V_F$ at which it was cooled.

\begin{figure*}[!]
\label{O2-dVdI}
\center{\includegraphics[width=14.5cm]{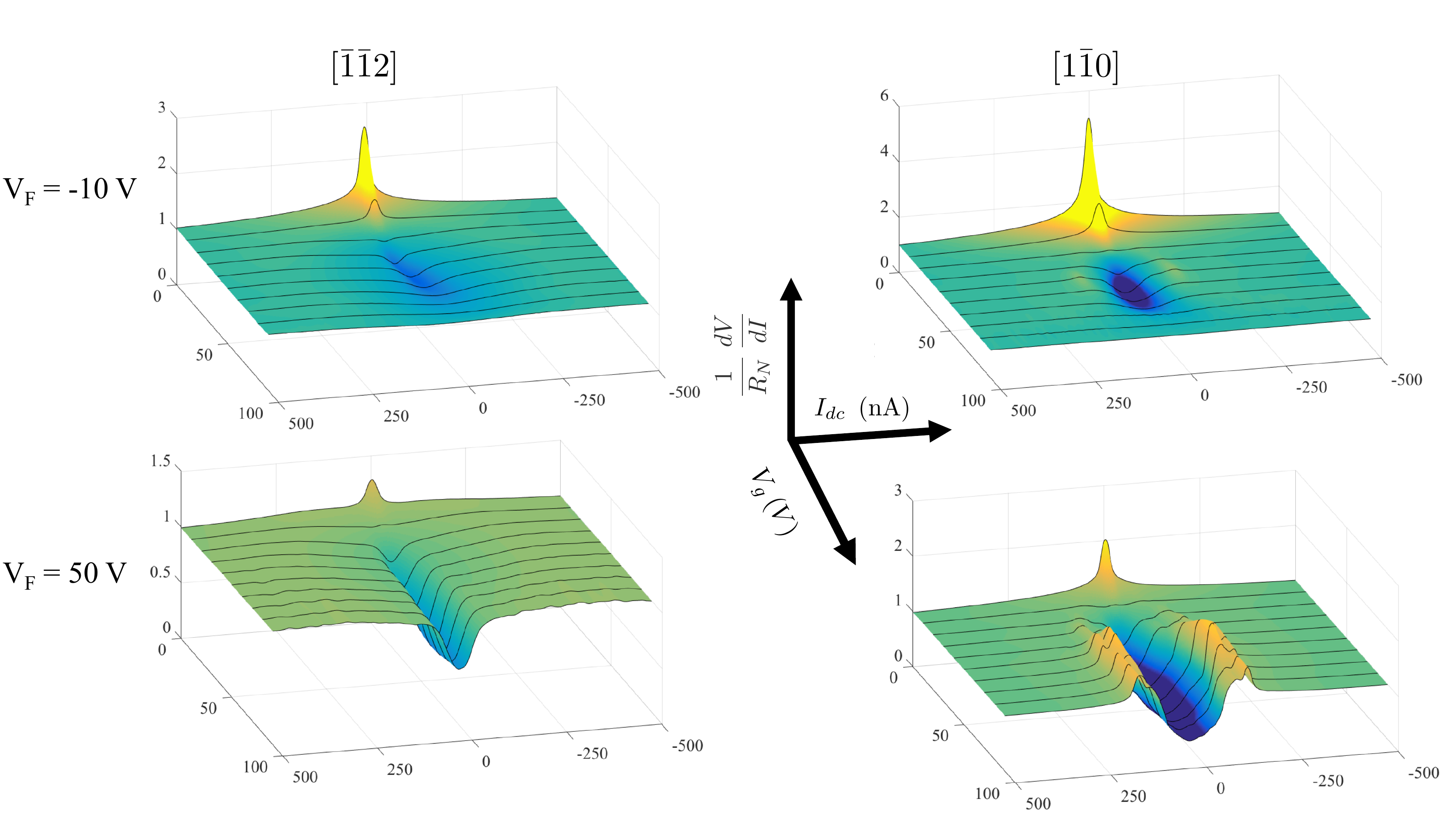}}
\caption{\textbf{Differential resistance of the O$_2$ annealed samples at 30 mK, for $V_F=-10$ V and $V_F=50$ V}.  While the O$_2$ annealed samples do not go fully superconducting, there is clear evidence for incipient superconductivity in the current-voltage characteristics, even for $V_F=-10$ V, for which the resistance increases with decreasing temperature (Fig. 1(b)).  As with the Ar/H$_2$ annealed samples (Fig. 2), the superconducting characteristics change with $V_g$, a consequence of electrostatic doping, but also are a function of the voltage $V_F$ at which they are cooled.  For the O$_2$ annealed samples, the superconducting characteristics are also highly anisotropic, being very different along the two crystal directions, as has been reported earlier for other transport properties \cite{davis2,davis1}.  Clear evidence for a maximum in superconducting properties as a function of $V_g$ (the superconducting `dome') can also be seen.} 
\end{figure*} 

Figure 3 shows similar data for the O$_2$ annealed samples.  Here, the samples do not go completely superconducting, although a dip in the differential resistance can clearly be seen over a certain range of $V_g$, mimicking the dome observed in (001) LAO/STO samples \cite{caviglia,dikin}.  Even so, the freezing effect is quite clear: in both crystal directions, the drop in differential resistance near zero bias along both crystal directions is much larger for $V_F=50$ V than for $V_F=-10$ V. We note that the O$_2$ annealed samples are also strongly anisotropic, which can be seen most strikingly in the data in Fig. \ref{O2-dVdI} for $V_F=50$ V.  (The anistropy in the Ar/H$_2$ annealed samples is much weaker, but still present.) 

A different representation of the data for the Ar/H$_2$ annealed samples in Fig. 2 is given in Figs. 4(a) and (b), which show $I_c$ and the normal state resistance $R_N$ as a function of $V_g$ for the Ar/H$_2$ annealed samples.  In general, $R_N$ increases rapidly with decreasing $V_g$, as expected, but the sharp upturn in $R_N$ occurs at larger $V_g$ for $V_F=-10$ V in comparison to $V_F=50$ V. $I_c$ shows a broad maximum, in effect, a saturation, at positive $V_g$, occurring at around $V_g \sim 0$ V for  
$V_F=50$ V, and $V_g \geq 100$ V for $V_F=-10$ V.  However, the $I_c's$ for $V_F=-10$ V are approximately a factor of 2 smaller than those for $V_F=50$ V.  Thus, setting a gate voltage $V_F=-10$ V as the sample is cooled through its transition results in lower critical currents  over the entire gate voltage range in comparison to $V_F=50$ V.  However, $T_c$ shows a different trend (Fig. 4(c)).  $T_c$ is lower for $V_F=-10$ V than for $V_F=50$ V for $V_g \leq -60$ V, but uniformly larger above this value.  This trend of an increase of $T_c$ coupled with a decrease in $I_c$ has also been observed in Al films\cite{nitt,abel}, and has been associated there with a modification with disorder of the attractive electron-phonon interaction responsible for superconductivity. \cite{deg,garland} 
\begin{figure}[h!]
\label{ICvsG}
\center{\includegraphics[width=6.5cm]{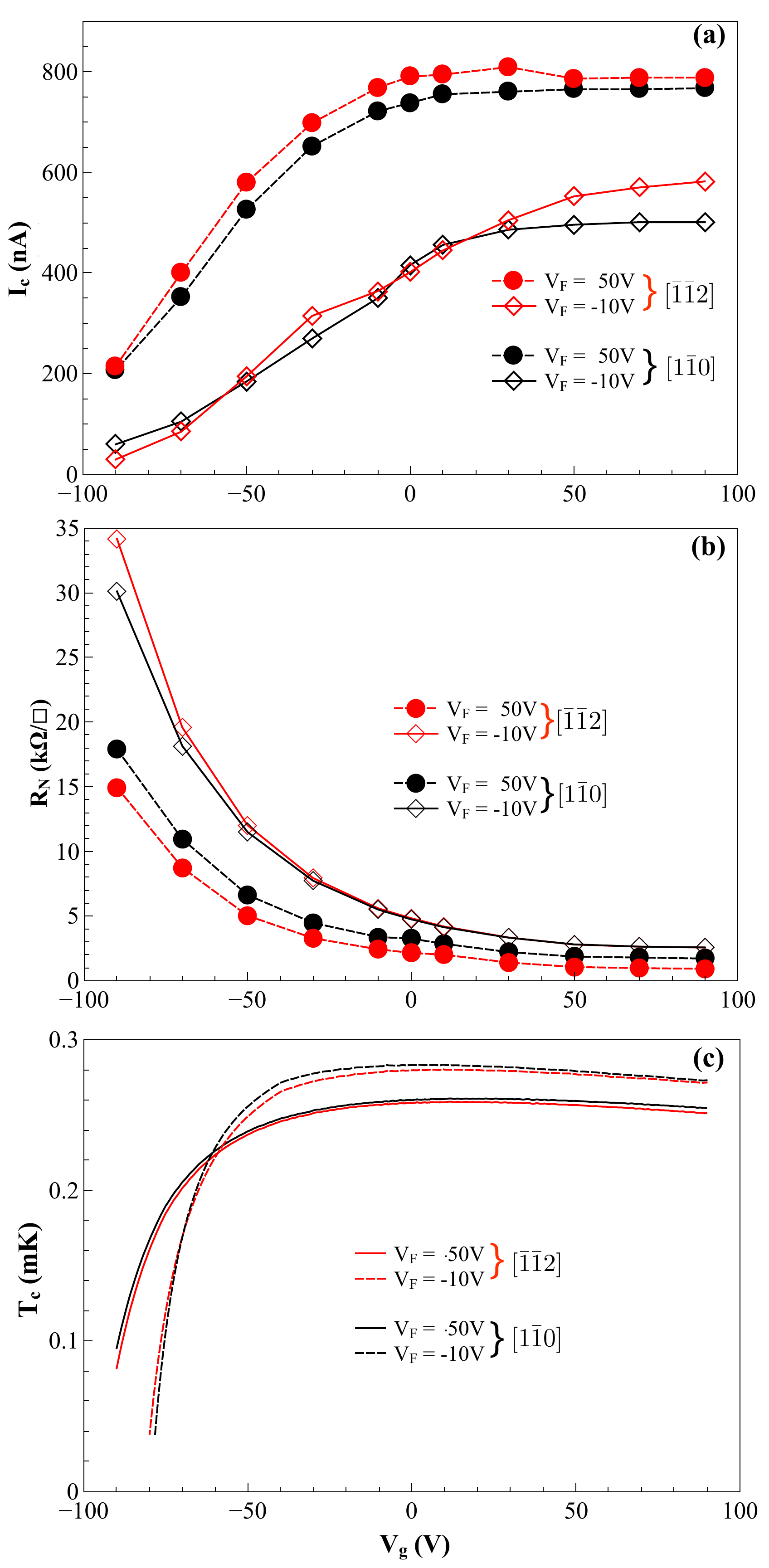}}
\caption{\textbf{Superconducting properties of the Ar/$H_2$ annealed samples for different freezing voltages.}  Critical current $I_c$ (a), normal state resistance (b) for the Ar/H$_2$ annealed samples for freezing voltages $V_F=-10$ V and $V_F=50$ V, for both the $[\bar{1}\bar{1}2]$ and $[1\bar{1}0]$ crystal directions, taken from Fig. 2.  Since the data show multiple peaks, $I_c$ is defined to be the value of $I_{dc}$ at which the slope of $dV/dI$ vs $I_{dc}$ is a maximum ($d^3V/dI^3 = 0$).  $R_N$ is defined as the value of $dV/dI$ at $I_{dc} \pm 1$ $\mu$A.  (c)  Critical temperature $T_c$ as a function of $V_g$, measured using feedback techniques by biasing the device at the foot of the superconducting transition while sweeping $V_g$ (see \emph{Methods}).}
\end{figure} 

From these data, it is evident that applying a back-gate voltage $V_F$ as the sample is cooled through its transition freezes in a specific configuration that influences the normal and superconducting state properties at very low temperatures, even if $V_g$ is subsequently swept. The configuration is stable in time over a scale of weeks, and robust against changes in $I_{dc}$ and magnetic field in addition to $V_g$, so long as the temperature is not raised above a certain freezing temperature $T_F$.  In order to determine $T_F$, we cooled the devices down from from 4.4 K to a target temperature with gate voltages of either $V_F=-10$ V or $V_F=50$ V applied, and then measured the resistance as a function of $V_g$.  After each resistance measurement, the sample was warmed to 4.4 K before another measurement was made.  These traces for 4 target temperatures of 30 mK, 470 mK, 675 mK and 1.2 K are shown in Fig. 5.  For simplicity, we have shown data for only the Ar/H$_2$ annealed devices in the $[\bar{1}\bar{1}2]$ crystal direction:  Data for the other crystal direction are shown in the Supplementary Information.  At 1.2 K and 675 mK, there is only a small difference in the resistance as a function of $V_g$.  At 470 mK, the difference is much larger, almost a factor of 5 at negative $V_g$.  Finally, at 30 mK, the difference is almost an order of magnitude at negative $V_g$ (the sample is superconducting for $V_g>-40$ V at this temperature).  Thus the temperature at which the configuration is frozen in is $T_F\sim 1$ K.

\begin{figure}[h!]
\label{VFCooldown}
\center{\includegraphics[width=6.5cm]{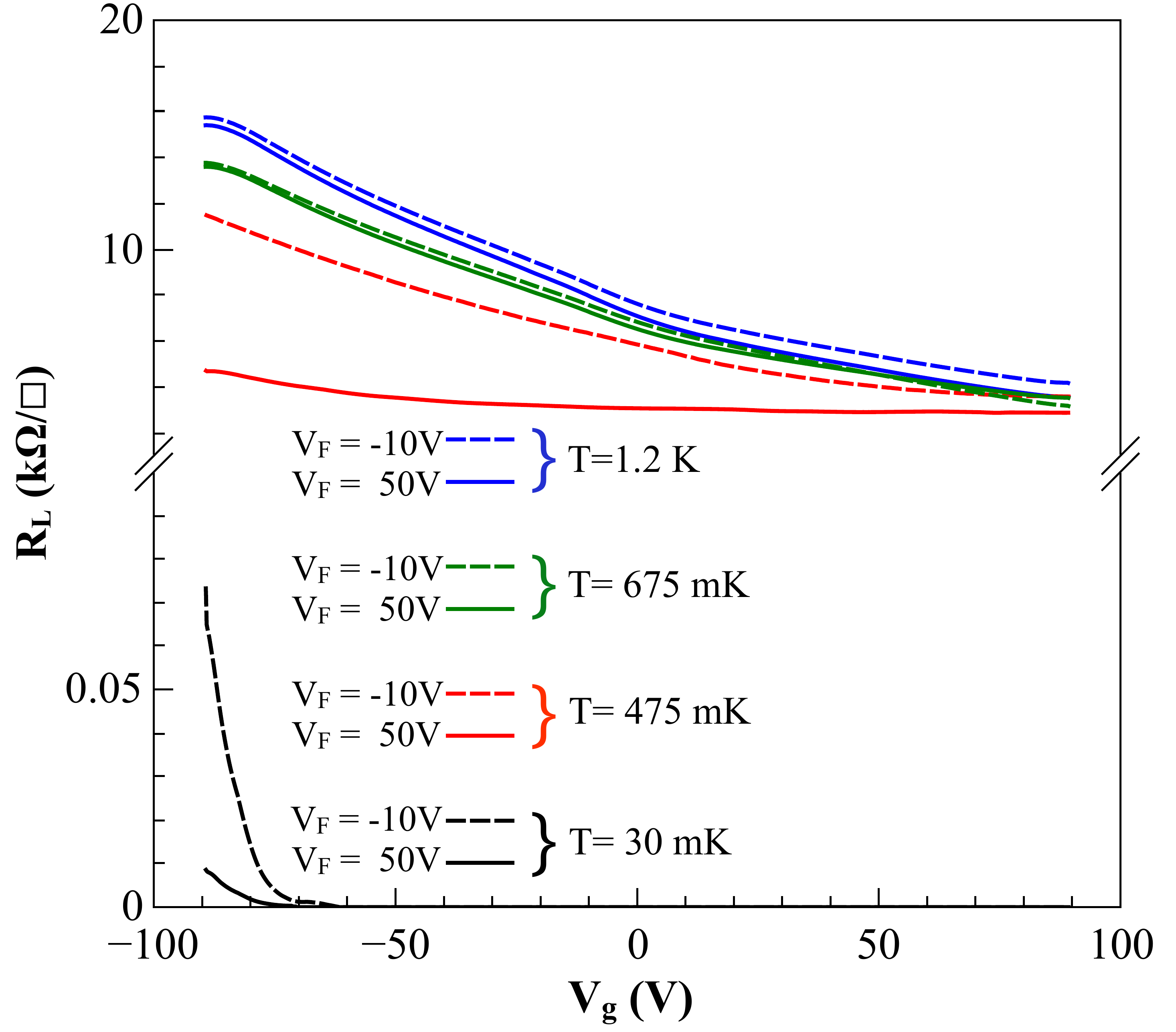}}
\caption{\textbf{Onset of the freezing transition.}  Resistance of the Ar/H$_2$ annealed samples in the $[\bar{1}\bar{1}2]$ crystal orientation as a function of $V_g$ after cooling from 4.4 K to the target temperatures noted with gate voltages $V_F=-10$ V and $V_F=50$ V applied.  As the resistance is hysteretic with $V_g$, the average of the upsweep and downsweep traces as a function of $V_g$ is shown in each case.  Differences between the curves for the two freezing temperatures show up only below 600-700  mK, showing that the freezing temperature $T_F$ is of order $\sim 0.6$ K.}
\end{figure} 

We now come to the question of the origin of this effect.  In (001) interface samples, the resistance at low temperature depends on the history of how $V_g$ is swept after first cooling the sample down from room temperature, eventually settling into a hysteretic but reproducible behavior as a function of $V_g$.  This has been ascribed to an irreversible change in the occupation of charge states near the interface that can only be reset by warming to room temperature, indicating that the relevant energy scales for population or depopulation of these charge states are of that order ($\sim$ 25 meV).  We also observe a similar initial, irreproducible, history dependence of the properties in the (111) interface samples on first cooling down from room temperature, after which the properties as a function of $V_g$ become hysteretic but reproducible.  However, the freezing effects in the (111) interface samples reported here are different, in that different values and $V_g$ dependences of $I_C$, $R_N$ and $T_C$ can be stabilized at very low temperature depending on the freezing voltage $V_F$ without introducing new irreproducible dependences at these temperatures.  Furthermore, these frozen state sets/resets at a much lower energy scale, $\sim 1$K.  It may be possible that the freezing effect is associated with other trapped states with much lower energy scales.  From Fig. 4, it appears that shifting the $V_F=50$ V curves towards positive $V_g$ would roughly align the curves for the two different freezing temperatures, although the shifts would be different for $R_N$, $I_c$, and $T_c$.  However, it is clear that even with shifting one of the curves with $V_g$, the superconducting properties cannot be made to align, as the saturation values of $I_c$ and $R_N$ for large $V_g$ for the two freezing temperatures are quite different.  Consequently, it is clear that the effects that we observe are not solely due to residual electric fields due to trapped charges, at least with regard to the superconducting properties: the effects we observe, particularly the increase in $T_c$ for $V_F=-10$ V, similar to what is observed in Al films, suggests that increased disorder due to the trapped charges may be important. If the freezing of charged trap states is indeed responsible for the freezing effect we observe, at 1K, one would need to understand the very low energy scales of these states.  Uncovering the origin of these low energy scales will require further experimental and theoretical work.

\section*{Methods }The 20 monolayer (ML) (111) LAO/STO interface samples reported on in this study were prepared by pulsed laser deposition using a KrF laser ($\lambda$ = 248 nm).  A LAO single crystal target was used for deposition, and the laser repetition was kept at 1 Hz, laser fluence at 1.8 J/cm$^2$, growth temperature at 650 C, and oxygen pressure at 1 mTorr. The deposition was monitored via \textit{in situ} reflection high energy electron diffraction (RHEED). Hall bars were then fabricated using photolithography to define an etch mask to protect the Hall bars during the subsequent argon ion milling.  The argon ion milling etched the unprotected areas to the bare STO, leaving the LAO on top of the Hall bars.  A final photolithography step deposited a Au film on the electrical contacts to enable visual location of the contacts for wire-bonding.  Additionally, Au was deposited on the etched bare STO, which enabled us to to confirm that the bare STO did not become conducting after the ion milling process or subsequent annealing steps.

The crystal orientation of the Hall bars shown in inset of Fig. 1(a) of the manuscript were determined via x-ray diffraction utilizing a Photonic Science Laue X-ray camera. The sample was mounted on a goniometer stage for precise angular alignment. The images were fit using the commercial PSL software provided with the camera, which allowed simultaneous determination of both Hall bar orientations on the substrate, as marked in the inset of Fig. 1(a) of the main manuscript. The samples were then subjected to post growth surface treatments to change the oxygen vacancy concentration at the interface. Four of the devices were subjected to O$_2$ annealing and other six devices were subjected to Ar/H$_2$ annealing. The exact annealing processes are described in Ref. \citenum{davis1}. The samples were cooled to millikelvin temperatures using both an Oxford MX100 and Kelvinox 300 dilution refrigerator. To measure the transport characteristics of the samples we utilized custom home-built amplifiers and high impedance current sources in conjunction with traditional lock-in techniques, allowing for measurement of samples impedances of up to a G$\Omega$. Further information about the feedback measurements used to measure $T_c$ continuously as a function of $V_g$ can be found in the Supplementary information. 

	


\end{document}